\documentclass[10pt]{iopart}
\usepackage{iopams}  
\expandafter\let\csname equation*\endcsname=\relax
\expandafter\let\csname endequation*\endcsname=\relaxD
\usepackage{epsf,graphicx,graphics}
\usepackage{amsmath}
\usepackage{latexsym,amssymb}
\usepackage{setspace,cite}
\usepackage{a4}
\usepackage[left=2cm,right=2cm,top=2cm,bottom=3cm]{geometry}

\begin{document}

\newtheorem{thr}{Theorem}
\newtheorem{defn}[thr]{Definition}
\newtheorem{lem}[thr]{Lemma}
\newtheorem{prop}[thr]{Proposition}
\newtheorem{cor}[thr]{Corollary}

\newcommand{\disfrac}[2]{\displaystyle{\frac{#1}{#2}}}

\newcommand{\con}[3]{\Gamma^{#1}_{#2 #3}}

\newcommand{\uone}{\mathfrak{u}(1)}
\newcommand{\essu}{\mathfrak{su}(2)}
\newcommand{\sun}{\mathfrak{su}(N)}

\newcommand{\bv}[1]{\underline{\mbox{\boldmath$#1$}}}

\newcommand{\rar}{\rightarrow}
\newcommand{\slim}{\sum\limits}

\newcommand{\drs}{\partial_{r_*}}
\newcommand{\drss}{\partial_{r_*}^2}

\newcommand{\R}{\mathbb{R}}

\newcommand{\mc}[1]{\mathcal{#1}}
\newcommand{\mk}[1]{\mathfrak{#1}}

\newcommand{\tbf}[1]{\textbf{#1}}

\newcommand{\newpar}{\\ \\}

\renewcommand{\thefootnote}{\arabic{footnote}}

\title[Stable topological EYM black holes with $\Lambda<0$]{Stable topological hairy black holes in $\sun$ EYM theory with $\Lambda<0$}

\author{J. Erik Baxter}

\address{c/o Faculty of ACES,
Sheffield Hallam University,
Howard Street,
Sheffield, 
South Yorkshire S1 1WB}
\ead{e.baxter@shu.ac.uk}

\begin{abstract}

We investigate the linear stability of topological black hole solutions to four-dimensional $\sun$ Einstein-Yang-Mills theory with a negative cosmological constant. We here extend recent results in the field which prove the existence of hairy black hole solutions to such equations \cite{baxter_existence_2015}, and the stability of their spherically symmetric analogues \cite{baxter_stability_2015}. We find the analysis in \cite{baxter_stability_2015} carries over very similarly, with some important differences in the final stages. Nevertheless, we establish the existence of non-trivial solutions stable under linear perturbations, in a sufficiently small neighbourhood of some existing trivial solutions; in fact, stable topological solutions turn out to be likely more abundant in the parameter space than their spherically symmetric analogues.

\end{abstract}

\vspace{2pc}

\section{Introduction}

Recent work has shed much light on hairy black holes and solitons -- that is, solutions to four-dimensional Einstein-Yang-Mills (EYM) theory. From the first solutions discovered in four-dimensional $\essu$ asymptotically flat space by Bartnik and McKinnon \cite{bartnik_particle-like_1988} in the soliton case and Bizon \cite{bizon_colored_1990} in the black hole case, interest has been high in investigating many generalisations of their work, for instance: to spaces of higher dimension or alternate topologies \cite{van_der_bij_new_2002, lemos_rotating_1996, cai_black_1996}; to `dyonic' solutions, characterised by a non-zero electric gauge sector \cite{nolan_existence_2012, radu_static_2002, radu_static_2004} (which only exist in spaces where $\Lambda<0$); and to extensions of the gauge group to $\sun$ or to abitrary compact gauge group \cite{kunzle_sun-einstein-yang-mills_1991, oliynyk_local_2002}.

Of central importance to the physicality of these solutions is the question of their stability. We may prove that certain solutions exist, but if we are to prove their \textit{longevity} in the universe, they must be stable under linear perturbations at the very least; otherwise we cannot be sure that if disturbed, the black hole hair will not radiate away or collapse inwards.

The first cases which were considered were all for asymptotically flat space. However, it was proven by Straumann and Zhou in \cite{straumann_instability_1990} that the original solutions discovered in $\essu$ were unstable under linear spherically symmetric perturbations. It was proven that the number of unstable modes of the $n^{th}$ solution (indexed by $n$, the number of zeroes (`nodes') that the gauge field function $\omega$ possesses) is $2n$, with $n$ in each of the two decoupled `sectors' (see later for details) and that for $\Lambda=0$, the gauge field must possess at least one zero \cite{lavrelashvili_remark_1995, volkov_number_1995, mavromatos_aspects_1996}. Zhou performed non-linear analysis on solitons in $\mathfrak{su}(2)$ and again showed them to be unstable, in that either hair radiates to infinity or else collapses to produce a Schwarzschild black hole \cite{zhou_instability_1990}. Finally, Brodbeck and Straumann proved instability for arbitrary gauge group in the asymptotically flat case \cite{brodbeck_instability_1996}. For a comprehensive review of stability in these cases, see \cite{volkov_gravitating_1999}.

It also makes physical sense that stability is directly related to the sign of the cosmological constant $\Lambda$ -- specifically, that solutions are only expected to be stable for $\Lambda<0$. This is a consequence of the `balancing act' between the repulsive forces set up by the gauge field and the attractive force of gravity; in the open geometry associated with $\Lambda\geq 0$, hair is free to radiate away to infinity if disturbed. However, $\Lambda<0$ corresponds to a closed geometry, meaning that the black hole hair is `trapped'. This is reflected in the nature of the solution space in each case: it is known that, in general, asymptotically flat hairy solutions occur only at discrete values of the parameters concerned, in other words their solution set does not form a continuum \cite{breitenlohner_static_1994, smoller_existence_1993, smoller_existence_1993-1, smoller_smooth_1991}, whereas for asymptotically anti-de Sitter (adS) space, solutions occur in continuous ranges of the parameter spaces and so are more plentiful \cite{winstanley_existence_1999, bjoraker_monopoles_2000, bjoraker_stable_2000, breitenlohner_non-abelian_2004}. This gives us hope that small perturbations in existing solutions will simply yield new solutions `not too far' from the original, and hence these solutions may be stable.

Many solutions have been found for $\Lambda<0$ in a variety of cases: see \cite{winstanley_classical_2009} for a recent review. The stability of some $\mathfrak{su}(2)$ solutions in asymptotically adS space is proven for $|\Lambda|\rightarrow\infty$ in \cite{winstanley_linear_2002, sarbach_linear_2001}. In addition, recent work \cite{baxter_stability_2015} shows stability for some solutions in spherically symmetric $\sun$ EYM theory. Given that we have recently established the existence of their topological analogues \cite{baxter_existence_2015}, it seemed natural to extend the work to cover a stability analysis and thus fill a gap in the research. Upon examination, it transpires that the analysis of the topological case differs from the spherical case in some small but vital details, which bear future consideration.

The purpose of this work is to examine those details, and so because of this we direct the reader towards \cite{baxter_stability_2015} and note that since the basic analysis is so similar, we shall only concentrate here on the cases for which $k\neq 1$ and focus solely on where the cases differ. We emphasise, this work is written to be read alongside \cite{baxter_stability_2015} as we rely heavily upon it. The outline of this paper is as follows. We begin by outlining all relevant formulae and ans\"{a}tze. We briefly review the static field equations, boundary conditions and trivial solutions, and move quickly on to the problem of stability. We find again that our choice to consider purely magnetic solutions allows the solution space to decouple into two sectors, which we take in turn -- we call these the `sphaleronic' and `gravitational' sectors. We find that the topology of the manifold makes little difference until the end of the analysis, when it becomes important: nevertheless we are able to use a theorem of Amann and Quittner's \cite{amann_nodal_1995} to prove the linear stability of non-trivial solutions in a small neighbourhood of existing trivial solutions, and in addition to this, we find that stable solutions appear to be more plentiful in the strictly topological case compared to the spherically symmetric case.
 
\section{Topological $\mathfrak{su}(N)$ Einstein-Yang-Mills theory}

In this section we give the general mathematical background we shall need to describe and model topological black holes. By way of review, we will also show the equilibrium field equations, boundary conditions and embedded solutions.

\subsection{Ans\"{a}tze}

The action used for the four-dimensional $\mathfrak{su}(N)$ Einstein-Yang-Mills (EYM) theory, with the cosmological constant $\Lambda<0$, is
\begin{equation}
S_{EYM}=\frac{1}{2}\int d^{4}x\sqrt{-g}[R-2\Lambda-\mbox{Tr}F_{\mu\nu}F^{\mu\nu}],
\end{equation}
where $R$ is the Ricci scalar, $F_{\mu\nu}$ is the non-Abelian gauge field and Tr signifies the Lie algebra trace. Throughout the paper the metric has signature $(-, +, +, +)$ and we use units in which $4\pi G=1=c$. 

Varying the action gives the field equations
\begin{equation}\label{origFE}
\begin{split}
2T_{\mu\nu}&=R_{\mu\nu}-\frac{1}{2}g_{\mu\nu}R + \Lambda g_{\mu\nu},\\
0&=\nabla_\lambda F^\lambda_{\:\:\mu}+[A_\lambda,F^\lambda_{\:\:\mu}],\\
\end{split}
\end{equation}
where the Yang-Mills (YM) stress-energy tensor $T_{\mu\nu}$ and the field strength tensor $F_{\mu\nu}$ are given by
\begin{equation}\label{origYMseten}
\begin{array}{ccc}
T_{\mu\nu}=\mbox{Tr}\left[F_{\mu\lambda}F_\nu^\lambda-\frac{1}{4}g_{\mu\nu}F_{\lambda\sigma}F^{\lambda\sigma}\right],&\quad\quad&F_{\mu\nu}=\partial_\mu A_\nu-\partial_\nu A_\mu+i[A_\mu,A_\nu],
\end{array}
\end{equation}
and we have rescaled so that the gauge coupling constant $g=1$. Note also that we have employed the usual Einstein summation convention where it is understood that summation occurs over repeated indices. 

In this paper we are interested in static, topological black hole solutions of the field equations \eqref{origFE}, specifically for spaces regularly foliated by 2D (spacelike) hypersurfaces of constant and unit- or zero-magnitude Gaussian curvature $k$, and hence (following \cite{baxter_existence_2015}) we write the metric in standard Schwarzschild co-ordinates as
\begin{equation}\label{topmet}
ds^2=-\mu S^2dt^2+\mu^{-1}dr^2+r^2d\theta^2+r^2f^2_k(\theta)d\phi^2,
\end{equation}
where $\mu$ and $S$ depend on $t$ and $r$ alone. For convenience, we may take 
\begin{equation}
\mu(t,r)=k-\frac{2m(t,r)}{r}-\frac{\Lambda r^2}{3}.
\end{equation}
%
%
%
The function $f_k(\theta)$ is used to endow the metric with the correct topology in the angular portion, and depends on our value of $k$ such that
\begin{equation}\label{fkdef}
f_k(\theta) = \left\{ \begin{array}{ll}
\sin\theta & \mbox{for } k=1, \\
\theta & \mbox{for } k=0, \\
\sinh\theta & \mbox{for } k=-1. \end{array}\right.
\end{equation}
It should stressed that we will solely be concerned with the non-spherically symmetric cases, i.e. $k=0,-1$. 
In \cite{baxter_existence_2015}, we used Wang's theorem \cite{wang_invariant_1958} to construct the following topological gauge potential:
\begin{equation}\label{GenGP}
A=\mathcal{A}\,dt+\mathcal{B}\,dr+\frac{1}{2}(C-C^\dagger)d\theta-\frac{i}{2}\left[(C+C^\dagger)f_k(\theta)+D\frac{df_k}{d\theta}\right]d\phi.
\end{equation}
In the above $\mathcal{A}, \mathcal{B}$, $D$ are all ($N\times N$) traceless, diagonal matrices, with $\mathcal{A}$ and $\mathcal{B}$ imaginary, $D$ real; $C$ is a complex ($N\times N$) upper-triangular matrix with non-zero entries only immediately above the main diagonal; and $C^\dagger$ is the Hermitian conjugate of $C$. (See \cite{baxter_existence_2015} for details.)
In comparison with (\tbf{2.8})\footnote{For brevity, all section numbers etc. which are in \textbf{bold} will be understood to refer to appropriate sections of \cite{baxter_stability_2015}; equations from \cite{baxter_stability_2015} will be simply given in the form (\tbf{X.XX}).}, the Einstein tensor $G_{\mu\nu}$ that we derive from the metric ansatz has components given by\footnote{Note that throughout this work, we denote $\dot{}$ to mean $\partial/\partial t$ and $^\prime$ to mean $\partial/\partial r$.}:
\begin{equation}\label{tensG}
G_{tt}=-\frac{\mu S^2}{r^2}\left(\mu'r-k+\mu\right), \quad G_{tr}=-\frac{\dot{\mu}}{\mu r}, \quad G_{rr}=\frac{1}{\mu Sr^2}\left(\mu'Sr+2S'\mu r-kS+\mu S\right).
\end{equation}
It is unnecessary to state the angular components of the Einstein equations since if (\ref{tensG}) hold, then the other components do automatically due to the Bianchi identities. We also need not state the components of $F_{\mu\nu}$ since three of the components of the field strength tensor ($F_{tr}$, $F_{t\theta}$ and $F_{r\theta}$) are the same as in (\tbf{2.10}) of \cite{baxter_stability_2015}, and three are slightly different: in $F_{t\phi}$, $F_{r\phi}$ and $F_{\theta\phi}$, the `$\sin\theta$' is replaced by $f_k(\theta)$.

Finally, we comment on the equilibrium field equations, the boundary conditions, and the embedded solutions here. The static field equations and symmetries are the same as (\tbf{2.17 - 2.22}) in \cite{baxter_stability_2015} but for the following changes:

\begin{itemize}
\item The function $\bar{\Pi}$ (\tbf{2.18}) is altered to $\bar{\Pi}=\disfrac{1}{4r^4}\slim_{j=1}^N\left(\left(\bar{\omega}_j^2-\bar{\omega}_{j-1}^2-k(N+1-2j)\right)^2\right)$; 
\item The function $W_j$ (\tbf{2.20}) becomes $W_{j,k}$:
\begin{equation}\label{Wjk}
W_{j,k}=k-\omega_j^2+\frac{1}{2}\left(\omega_{j+1}^2+\omega_{j-1}^2\right).
\end{equation}
\end{itemize}
The stated boundary conditions (\tbf{2.23 - 2.26}) are identical but for the alteration of (\tbf{2.24}) to $\bar{m}(r_h)=\disfrac{kr_h}{2}-\disfrac{\Lambda r_h^3}{6}$. There will be some changes in the higher order parameters which involve $k$ but in this paper, as in \cite{baxter_stability_2015}, we are only interested in the leading order behaviour of the field variables. As for the embedded solutions, we note that the SAdS solution (\tbf{2.27 - 2.28}) is unavailable for $k\neq1$, but the other two solutions remain: the Reissner-N\"{o}rdstrom-Topological-anti-de-Sitter (RNTadS) solution (\tbf{2.29 - 2.30}), which has an altered magnetic charge (\tbf{2.31}) given by
\begin{equation}
Q_M=\frac{k^2N(N^2-1)}{6};
\end{equation}
and the embedded $\essu$ solution (\tbf{2.32}), defined by letting $\bar{\omega}_j=\sqrt{j(N-j)}\bar{\omega}$.
The only other detail which it may be important to state is that the bound given by requiring $m'(r_h)>0$ leads to a minimum event horizon radius for $k=-1$, given by
\begin{equation}
|r_h|\geq\sqrt{|\Lambda|(2m'(r_h)+1)},\,\,\mbox{ i.e. }\,\,|r_h|>\left(|\Lambda|-2r_h^2\bar{\Pi}(r_h)\right)^{-\frac{1}{2}},
\end{equation}
which in turn implies a minimum bound on $|\Lambda|$, generalising a result in \cite{van_der_bij_new_2002}:
\begin{equation}
|\Lambda|>\frac{1}{r_h^2}\left(1+2r_h^2\bar{\Pi}(r_h)\right).
\end{equation}

\section{Perturbation equations}

We wish to examine the behaviour of solutions to the field equations under linear perturbations, so all field variables ($\mu$, $m$, $S$, $\omega_j$, $\alpha_j$, $\beta_j$, $\gamma_j$) are rewritten the form
\begin{equation}
f(t,r)=\bar{f}(r)+\delta f(t,r)
\end{equation}
(where from now on, a bar $\bar{ }$ represents the equilibrium solution variables, which are dependent on $r$ alone, and where $\bar{\alpha}_j=\bar{\gamma}_j\equiv0$), and the resulting perturbation equations are obtained. It should be noted that in our case, the perturbations will of course not be spherically symmetric, but will share the symmetry (planar or hyperbolic) of the angular portion of the topological metric \eqref{topmet}. We are only interested in purely magnetic solutions; therefore we have a zero electric gauge field and we set the perturbations $\delta{\alpha_j}\equiv0$. In this regime, the equations decouple into two sectors: the \textit{sphaleronic sector}\footnote{This sector is so named because the original $\essu$ EYM solitons \cite{bartnik_particle-like_1988} and black holes \cite{bizon_colored_1990} in asymptotically flat space possess instabilities in this sector \cite{volkov_number_1995, lavrelashvili_remark_1995} analogous to the unstable mode of the Yang-Mills-Higgs sphaleron \cite{yaffe_static_1989}.}, containing only the YM perturbations $\delta\beta_j$ and $\delta\gamma_j$; and the \textit{gravitational sector}, containing only the metric perturbations $\delta\mu$ and $\delta S$ and the YM perturbations $\delta\omega_j$. Thus, these two sectors are considered separately. 

A final point to make before we do this is that Sections \tbf{2.3}, \tbf{3.1 - 3.4}, \tbf{3.6}, \tbf{4.1} and \tbf{4.2} carry across almost identically, with two main (and small) differences, both of which are the same as the alterations in the field equations (see the list following equation \eqref{tensG}); the result of which is that only terms involving $W_{j,k}$ actually make a difference to the final analysis. As we will show though this does make a fairly significant difference to the results.

\subsection{Sphaleronic sector}

The basic idea here was to use co-ordinate transforms to organise the variables $\delta\beta_j$ and $\delta\gamma_j$ into a single vector $\bv{\Psi}$ and thus to state their second order differential equations in a self-adjoint hyperbolic form (\tbf{3.19}):
\begin{equation}
\mathcal{U}\bv{\Psi}=-\ddot{\bv{\Psi}}
\end{equation}
Then, for time-periodic perturbations, i.e. for which $\bv{\Psi}(t,r)=\mbox{Re}\left(e^{i\sigma t}\tilde{\bv{\Psi}}(r)\right)$, the perturbations will take the form 
\begin{equation}
\mathcal{U}\tilde{\bv{\Psi}}=\sigma^2\tilde{\bv{\Psi}},
\end{equation}
so that if $\mathcal{U}\geq0$ then the eigenvalues $\sigma^2\geq0$, so that $\sigma\in\mathbb{R}$, making $e^{i\sigma t}\tilde{\bv{\Psi}}(r)$ a time-periodic perturbation and thus stable. (If $\mc{U}=0$, we get no dynamics and the only perturbations correspond to gauge transformations -- hence we get trivial stability.)

We arranged the operator $\mc{U}$ into the form  $\mathcal{U}=\chi^\dagger\chi+\mathcal{V}-\mathcal{G}^\dagger h^2\mathcal{G}$, where $\chi$ and $\mc{V}$ are matrices in block-matrix form, $\mc{G}$ is the operator representing the Gauss constraint, and $h=\frac{\bar{S}\sqrt{\bar{\mu}}}{r}$. Due to the forms of the various terms involved, and the fact that the last term will vanish when applied to physical perturbations that satisfy the strong Gauss constraint, we ensured that $\mc{U}$ was hyperbolic, symmetric and real; and also positive, given only one constraint appearing in the matrix $\mc{V}$: that is,
\begin{equation}\label{Wleq0}
\mc{W}\leq0,
\end{equation}
for the matrix $\mc{W}$ (\tbf{3.15}). Following the analysis, this means the argument is structurally identical in our case; and in fact the only difference that survives the process is in $\mathcal{W}$, and hence the matrix element $\mathcal{V}_{22}$ (\tbf{3.39}). Therefore, the main difference in the analysis of Sections \tbf{2 - 4} is in the stability condition \eqref{Wleq0}, dealt with in Sections \tbf{3.5} and \tbf{4.3} -- the so-called \textit{special cases}. What we need to consider first is what \eqref{Wleq0} implies for those cases.

\subsubsection{Conditions for no sphaleronic sector instabilities}

Comparing our work with Section \tbf{3.1 - 3.4}, the only difference is in the matrix $\mc{W}$ (\tbf{3.15}), where in our case
\begin{equation}
\mathcal{W}=h^2\mbox{diag}\left(\bar{W}_{1,k},\bar{W}_{2,k},\ldots\right),
\end{equation}
for which $\bar{W}_{j,k}$ is the equilibrium part of $W_{j,k}$, defined in \eqref{Wjk}. Therefore for the matrix $\mathcal{W}$ to be non-positive we require $\bar{W}_{j,k}\leq0$, which gives us a set of inequalities that must be satisfied by the equilibrium solutions, for $j=1, ..., N-1$ -- these are the analogues of (\tbf{3.42}):
\begin{equation}\label{wkstabcon}
\bar{\omega}_j^2\geq k+\frac{1}{2}\left(\bar{\omega}^2_{j+1}+\bar{\omega}^2_{j-1}\right).
\end{equation}
The first thing to notice is that if we take $k=0,-1$, the right-hand side of \eqref{wkstabcon} is lower than if $k=1$, so na\"{i}vely we may think that stable solutions in these cases might be more abundant. Once again though, the challenge is to prove that we can find equilibrium solutions which conform to \eqref{wkstabcon} for all $r$.

\subsubsection{Special cases}

As in \cite{baxter_stability_2015}, we must next consider the embedded solutions discussed in Sections \tbf{3.5.2}, \tbf{3.5.3}.

\subsubsection*{The RNTadS solution}

Here we let $\bar{\omega}_j\equiv0$ for $j=1, ..., N-1$. We find that the perturbations reduce down considerably, to the point where the only allowed solution for bound state perturbations is the trivial solution $\bv{\Psi}=0$. Hence, there are no dynamics in the sphaleronic sector in this case, and the only bound state perturbations are pure gauge transformations. So the RNTadS solution is trivially stable here. This is identical to the $k=1$ case.

\subsubsection*{$\essu$ embedded solutions}

Here we let $\bar{\omega}_j=\sqrt{j(N-j)}$ for $j=1, ..., N-1$. In that case, 
\begin{equation}
\mc{W}=h^2(k-\bar{\omega}^2)\mc{I}_{N-1},
\end{equation}
which means that the condition $\mc{W}\leq 0$ implies
\begin{equation}\label{embedsucond}
\bar{\omega}^2\geq k
\end{equation}
for all $r$. Thus, all topological $\essu$ solutions immediately satisfy \eqref{embedsucond} for all functions $\bar{\omega}$. 

Finally, we note that since these embedded $\essu$ solutions satisfy \eqref{wkstabcon}, the analysis of Section \tbf{3.6} in \cite{baxter_stability_2015} carries straight over (using instead the analogous existence propositions from \cite{baxter_existence_2015} to construct the nearby solution); and hence we can still find non-trivial solutions which are sufficiently close to existing embedded solutions, within some small enough neighbourhood of the parameter space, and which possess no instabilities in the sphaleronic sector.

\subsection{Gravitational sector}

The situation is somewhat similar to the sphaleronic sector, though the analysis is far more resistant to simplification in the general case. We begin by defining a vector $\bv{\delta\omega}=(\delta\omega_1, \delta\omega_2, ... \delta\omega_{N-1})^T$. After manipulating the perturbation equations to make the problem more tractable, we showed that the gravitational sector is naturally self-adjoint. Thus we expressed the YM perturbations $\delta\omega_j$ in the following matrix form (\tbf{4.8}):
\begin{equation}
-\bv{\delta\ddot{\omega}}=-\drss\bv{\delta\omega}+\mc{M}\bv{\delta\omega}
\end{equation}
Therefore, for all time-periodic perturbations such that $\bv{\delta\omega}(t,r)=e^{i\sigma t}\bv{\delta\tilde{\omega}}(r)$, we have
\begin{equation}\label{genom}
\sigma^2\tilde{\bv{\delta\omega}}=-\drss\tilde{\bv{\delta\omega}}+\mathcal{M}\tilde{\bv{\delta\omega}}.
\end{equation}
Since \eqref{genom} is a a Schr\H{o}dinger-like equation, the non-negativity of $\mc{M}$ implies that $\sigma\in\R$ and hence there are no unstable modes in the gravitational sector; so we wish to prove in each case that $\mc{M}\geq0$.

As in the $k=1$ case, the matrix $\mc{M}$ (\tbf{4.9}) can be separated into entries that are on the main diagonal ($\mc{M}_{j,j}$), entries just above and below the main diagonal ($\mc{M}_{j,j+1}=\mc{M}_{j+1,j}$), and all other entries ($\mc{M}_{j,k}$), as follows:

\begin{eqnarray}\label{mplus}
\nonumber\mathcal{M}_{j,j}&=&-\frac{\bar{\mu}\bar{S}^2}{r^2}(\bar{W}_{j,k}-2\bar{\omega}^2_j)-\frac{4}{\bar{\mu}\bar{S}r}Q(\drs\bar{\omega}_j)^2-\frac{8\bar{S}}{r^3}(\drs\bar{\omega}_j)\bar{W}_{j,k}\bar{\omega}_j\\
\nonumber\mathcal{M}_{j,j+1}&=&-\frac{\bar{\mu}\bar{S}^2}{r^2}\bar{\omega}_j\bar{\omega}_{j+1}-\frac{4}{\bar{\mu}\bar{S}r}Q(\drs\bar{\omega}_j)(\drs\bar{\omega}_{j+1})-\frac{8\bar{S}}{r^3}\left(\bar{W}_{j,k}\bar{\omega}_j\drs\bar{\omega}_{j+1}+\bar{W}_{j+1,k}\bar{\omega}_{j+1}\drs\bar{\omega}_j\right)\\
\nonumber\mathcal{M}_{j,l}&=&-\frac{4}{\bar{\mu}\bar{S}r}Q(\drs\bar{\omega}_j)(\drs\bar{\omega_l})-\frac{8\bar{S}}{r^3}\left(\bar{W}_{j,k}\bar{\omega}_j\drs\bar{\omega}_l+\bar{W}_{l,k}\bar{\omega}_l\drs\bar{\omega}_j\right)\\
\end{eqnarray}
where $r_*$ is the `tortoise' co-ordinate introduced in \cite{baxter_stability_2015} (and e.g. \cite{winstanley_existence_1999}), i.e.
\begin{equation}
\frac{dr_*}{dr}=\frac{1}{\bar{\mu}\bar{S}}.
\end{equation}

\subsubsection{The RNTadS solution}
The first embedded solution we shall test is the RNTadS solution, which is found by letting $\bar{\omega}_j\equiv0$ for all $j$. Then we find that $\bar{W}_{j,k}\equiv k$, and the matrix $\mc{M}$ in \eqref{mplus} reduces right down to (in analogy with (\tbf{4.15}))
\begin{equation}
\mathcal{M}=-k\frac{\bar{\mu}\bar{S}^2}{r^2}\mathcal{I}.
\end{equation}

Let's take the 2 cases of $k$ which concern us:

\subsubsection*{$k=0$:}

Here, $\mc{M}$ is the zero matrix, and so the equation reduces to $\sigma^2\bv{\tilde{\delta\omega}}=-\drss\bv{\delta\omega}$. Since the right-hand side of this is still positive, then $\sigma^2>0$, giving $\sigma\in\mathbb{R}$; and therefore the RNTadS solution has a stable gravitational sector here.

\subsubsection*{$k=-1$:}

Now, the matrix $\mathcal{M}$ has positive eigenvalues, again meaning the right-hand side is positive giving $\sigma\in\mathbb{R}$, and therefore once more the RNTAdS solution has a stable gravitational sector. 

We note that both of these results are directly the opposite to the spherically symmetric case, and so we have result of some interest. It appears that we have a solution in which $\omega_j\equiv0$ for all $j$,  which as we have explained, we might not necessarily expect stability from -- however, the RNTadS solution is stable for $k\neq1$. We will return to this point in Section \ref{conclu}.

\subsubsection{$\mathfrak{su}(2)$ embedded solutions}

Embedding $\mathfrak{su}(N)$ in $\mathfrak{su}(2)$ means making the following identification:
\begin{equation}
\bar{\omega}_j=\bar{\omega}(r)\sqrt{j(N-j)}.
\end{equation}
We begin by rewriting the non-zero elements of the original $\mathcal{M}$ matrix using this identification. We note in this case that $\bar{W}_{j,k}=k-\bar{\omega}^2$. Then we find we can write the matrix $\mathcal{M}$ (\ref{mplus}) more conveniently by separating the matrix into three separate terms (\tbf{4.17 - 4.18}):
\begin{equation}
\mc{M}=\mc{N}_1+\mc{N}_2+\mc{N}_3.
\end{equation}
However, $\mc{N}_2$ and $\mc{N}_3$ are identical to those in \cite{baxter_stability_2015}, thus neither contain any reference to $k$ and so their positivity is assured; unconditionally for $\mc{N}_2$, and for $\mc{N}_3$ under the condition that $|\Lambda|\rar\infty$. Therefore we only need to examine $\mc{N}_1$:
\begin{equation}
\mathcal{N}_1=\frac{\bar{\mu}\bar{S}^2}{r^2}(\bar{\omega}^2-k)\mathcal{I}.
\end{equation}
We bear in mind that for stability, we want this part to be non-negative, in which case we have
\begin{equation}
\bar{\omega}^2\geq k.
\end{equation}
This is the same as the condition for stability in the sphaleronic sector; and it is guaranteed to be fulfilled for $k=0,-1$. Therefore $\mc{N}_1\geq0$, $\mathcal{M}$ is non-negative, and thus all $\essu$ embedded solutions possess no unstable nodes in the gravitational sector provided $|\Lambda|\rar\infty$. Comparing the $k=1$ case then, we see that this situation is similar but \textit{without} any conditions on $\omega_j$, meaning our region of stable solutions in the parameter space should in principle be larger than in the spherically symmetric space. Note that the regime of $|\Lambda|$ large was also used in \cite{van_der_bij_new_2002} to prove stability in the gravitational sector for topological $\essu$ solutions.

We briefly summarise our results so far. In the sphaleronic sector, the RNTadS solution has no dynamics; the $\essu$ embedded solution is unconditionally stable; and we can find nearby non-trivial solutions stable in the sector, i.e. they also fulfil the condition \eqref{wkstabcon}. Therefore, we only need to be concerned with the gravitational sector; in which the RNTadS solution is unconditionally stable and the $\essu$ embedded solution is stable for large $|\Lambda|$. All there is left to do is prove that non-trivial solutions exist which are stable in the gravitational sector, in a neighborhood of any $\essu$ embedded solutions. The rest of Section \tbf{4} employs a theorem by Amann and Quittner \cite{amann_nodal_1995} that we will refer to as the \textit{Nodal Theorem}.

\subsubsection{The Nodal Theorem}

This theorem (quoted in Section \tbf{4.4.1}) concerns the eigenvalue problem for a certain Schr\H{o}dinger-like linear differential operator, and determines the number of bound states of such an operator in terms of the nodes of the solution vectors of an auxiliary initial value problem (\tbf{4.26}). We note that Section \tbf{4.4} carries over virtually unchanged (ignoring Section \tbf{4.4.3} on solitons): in particular, equations (\tbf{4.28}) to (\tbf{4.35}) in Section (\tbf{4.4.2}) are identical to our case but for one occurrence of $k$ in $\mc{M}_{j,j}$ (\tbf{4.31}), which becomes 
\begin{equation}
\mc{M}_{j,j}=\frac{\Lambda}{3}\left[k-3\bar{\omega}_{j,\infty}+\frac{1}{2}\left(\bar{\omega}_{j-1,\infty}+\bar{\omega}_{j+1,\infty}\right)\right]+O(r^{-1}),
\end{equation} 
though this doesn't change the leading order of the expression and hence the application of the Nodal Theorem is unaffected. Thus, it can be proven that $\essu$ embedded solutions possess no unstable modes in the gravitational sector.

Finally it remains to prove that we can find non-trivial solutions in some neighbourhood of these embedded solutions that still satisfy the Nodal Theorem. This is dealt with in Section \tbf{4.5}. The basic idea is that for the $\essu$ embedded solutions we prove that the Nodal Theorem is fulfilled as long as the function defined by
\begin{equation}\label{frho}
\mk{F}(\rho)=\det\mk{U}_{\mk{c}}(\rho)
\end{equation}
has no zeroes over a particular range, where $\mk{U}_{\mk{c}}=[\bv{u}_1,\bv{u}_2,...,\bv{u}_{N-1}]$ (\tbf{4.25}) is the matrix of solution vectors to the eigenvalue problem (\tbf{4.26}), and the vectors $\bv{u}_j$ represent the gauge field perturbations $\delta\omega_j$. We then must apply the theorem to prove that we can find non-trivial solutions nearby the embedded $\essu$ solutions\footnote{We note that we cannot apply the Nodal Theorem to the RNTadS solution, even though it is stable, because the gauge field in that case is identically zero and thus the condition on $\mk{F}(\rho)$ is immediately violated.} which are stable in the sphaleronic sector and fulfil \eqref{frho}. Happily the argument in Section (\tbf{4.5}) carries over identically: as before we use an embedded $\essu$ solution, which is stable in both sectors given large $|\Lambda|$, and for which $\mk{F}(\rho)\neq0$; and we use analogous existence results from \cite{baxter_existence_2015}, which we use to establish analyticity of boundary conditions and particularly that solutions nearby the embedded solution will also have $\mk{F}(\rho)\neq0$ over the range of interest. 

At last, we deduce that if solutions stable in the sphaleronic sector exist in some neighbourhood $\mk{R}_S$ of stable embedded solutions, and solutions stable in the gravitational sector exist in some neighbourhood $\mk{R}_G$ of stable embedded solutions, then we can find solutions in $\mk{R}_S\cap\mk{R}_G$ which are stable in both sectors. 

\section{Conclusions}
\label{conclu}

In this paper we have proven the stability of some non-trivial black hole solutions to purely magnetic topological $\sun$ Einstein-Yang-Mills equations (with $\Lambda<0$) under linear perturbations which share the same topology as the angular portion of the metric \eqref{topmet}. Our focus has been to describe precisely how the change in topology affects the analysis carried out in \cite{baxter_stability_2015}.

First we applied a linear perturbation to the appropriate equilibrium field equations \cite{baxter_existence_2015}. We found that the purely magnetic system decouples into two sectors. The sphaleronic sector, after some manipulation, yielded a hyperbolic system, requiring a positive operator for stability. We deduced that the sphaleronic sector would have no dynamics for the RNTadS solution, and was unconditionally stable for $\essu$ embedded solutions, since the condition for stability for $\essu$ topological black holes \eqref{embedsucond} is guaranteed fulfilled. We then showed that genuine $\sun$ solutions could be found in some neighbourhood of these embedded solutions which have no instabilities in the sphaleronic sector, since the argument from \cite{baxter_stability_2015} carries over.

We then examined the gravitational sector. We showed that the metric perturbations could be eliminated and obtained a self-adjoint Schr\H{o}dinger-like matrix equation in $\delta\omega_j$, for which we once again required a positive definite matrix for stability. We based our approach on \cite{baxter_stability_2015}, using a theorem of Amann and Quittner's \cite{amann_nodal_1995} concerning the number of bound states of a Schr\H{o}dinger-like operator. We can again prove the existence of non-trivial solutions with no instabilities in the gravitational sector, in some neighbourhood of existing embedded $\essu$ solutions, in the regime of $|\Lambda|\rightarrow\infty$. Finally, we reasoned that there must exist some solutions in the intersection of the two neighbourhoods of the embedded solutions which possess no unstable modes in either sector.

The main result of this paper is the existence of solutions to purely magnetic, topological, asymptotically adS $\sun$ field equations, in some neighbourhood of known existing solutions, which are stable under linear perturbations in the limit of large $|\Lambda|$. We also note a slightly surprising result that came from our analysis of the RNTadS solution. As we mentioned, we expect (by analogy with the asymptotically flat case) that for $\Lambda<0$, at least some nodeless solutions will be stable, and therefore we suspected (by analogy with the $\essu$ spherical case \cite{winstanley_existence_1999}) that nodelessness may be a sufficient (but not necessary) condition for stability; and so in our existence proofs we have searched primarily for nodeless solutions as being more likely to be stable. Here we find examples of solutions for which nodelessness is not necessary for stability, seemingly due to the topology of the manifold. This discussion is outside the scope of this work but perhaps it would merit future study. 

Another possible way this work could be extended would be to use numerical methods to empirically verify the assertions we have made and to find some examples of stable topological solutions. Such an analysis was carried out for the spherically symmetric case \cite{baxter_stability_2015, baxter_abundant_2008}, where it was verified that as $|\Lambda|$ increased, the parameter space of solutions satisfying the stability condition in that case grew exponentially larger. Therefore two obvious questions to ask are what the space of solutions looks like in this case, and precisely how the abundance of the solutions is affected by the value of $|\Lambda|$ and the $N-1$ initial gauge field parameters -- perhaps some bounds could be established connecting these quantities. 

A further natural future direction of this work would be to revisit the ``No-hair'' conjecture as formulated by Bizon \cite{bizon_colored_1990}, in light of these results, which is stated as:
\begin{quote}
``Within a given matter model, a \textit{stable} black hole is uniquely determined by global charges.''
\end{quote}
In light of the `No-hair' conjecture, then, it can be seen that the stability of solutions is arguably more important an issue in the context of gravitational physics than their mere existence; but that is not the whole story, since there is the question of whether `global charges' can be defined in this case. In \cite{shepherd_characterizing_2012}, it is argued that spherically symmetric purely magnetic $\sun$ black holes can be characterised by a finite set of charges at infinity, and it would be of interest to know whether this was true for topological black holes as well, or whether the change in the entropy-temperature curves might cause a significant difference to the analysis.



\section*{References}

\vspace{0.3cm}

\appendix

%
%
%
\bibliography{Topological_stability_SU(N)}

\begin{thebibliography}{10}

\bibitem{baxter_existence_2015}
J.~E. Baxter.
\newblock {\em Gen. Rel. Grav.}, \textbf{47}:1829, 2015.

\bibitem{baxter_stability_2015}
J.~E. Baxter and E.~Winstanley.
\newblock {\em arXiv:1501.07541 [gr-qc, physics:hep-th]}, 2015.

\bibitem{bartnik_particle-like_1988}
R.~Bartnik and J.~McKinnon.
\newblock {\em Phys. Rev. Lett.}, \textbf{61}:141--144, 1988.

\bibitem{bizon_colored_1990}
P.~Bizon.
\newblock {\em Phys. Rev. Lett.}, \textbf{64}:2844--2847, 1990.

\bibitem{van_der_bij_new_2002}
J.~J. van~der Bij and E.~Radu.
\newblock {\em Phys. Lett. B}, \textbf{536}:107--113, 2002.

\bibitem{lemos_rotating_1996}
J.~P.~S. Lemos and V.~T. Zanchin.
\newblock {\em Phys. Rev. D}, \textbf{54}:3840--3853, 1996.

\bibitem{cai_black_1996}
R.-G. Cai and Y.-Z. Zhang.
\newblock {\em Phys. Rev. D}, \textbf{54}:4891--4898, 1996.

\bibitem{nolan_existence_2012}
B.~C. Nolan and E.~Winstanley.
\newblock {\em Class. Quant. Grav.}, \textbf{29}:235024, 2012.

\bibitem{radu_static_2002}
E.~Radu.
\newblock {\em Phys. Rev. D}, \textbf{65}:044005, 2002.

\bibitem{radu_static_2004}
E.~Radu and E.~Winstanley.
\newblock {\em Phys. Rev. D}, \textbf{70}:084023, 2004.

\bibitem{kunzle_sun-einstein-yang-mills_1991}
H.~P. Kunzle.
\newblock {\em Class. Quant. Grav.}, \textbf{8}:2283--2297, 1991.

\bibitem{oliynyk_local_2002}
T.~A. Oliynyk and H.~P. Kunzle.
\newblock {\em J. Math. Phys.}, \textbf{43}:2363--93, 2002.

\bibitem{straumann_instability_1990}
N.~Straumann and Z.~H. Zhou.
\newblock {\em Phys. Lett.}, \textbf{B237}:353, 1990.

\bibitem{lavrelashvili_remark_1995}
G.~V. Lavrelashvili and D.~Maison.
\newblock {\em Phys. Lett. B}, \textbf{343}:214--217, 1995.

\bibitem{volkov_number_1995}
M.~S. Volkov, O.~Brodbeck, G.~Lavrelashvili, and N.~Straumann.
\newblock {\em Phys. Lett. B}, \textbf{349}:438--42, 1995.

\bibitem{mavromatos_aspects_1996}
N.~Mavromatos and E.~Winstanley.
\newblock {\em Phys. Rev. D}, \textbf{53}:3190--214, 1996.

\bibitem{zhou_instability_1990}
Z.~H. Zhou.
\newblock {\em Phys. Lett. B.}, \textbf{237}:353, 1990.

\bibitem{brodbeck_instability_1996}
O.~Brodbeck and N.~Straumann.
\newblock {\em J. Math. Phys.}, \textbf{37}:1414--1433, 1996.

\bibitem{volkov_gravitating_1999}
M.~S. Volkov and D.~V. Gal'Tsov.
\newblock {\em Phys. Rept.}, \textbf{319}:1--83, 1999.

\bibitem{breitenlohner_static_1994}
P.~Breitenlohner, P.~Forgacs, and D.~Maison.
\newblock {\em Comm. Math. Phys.}, \textbf{163}:141--172, 1994.

\bibitem{smoller_existence_1993}
J.~A. Smoller, A.~G. Wasserman, and S.-T. Yau.
\newblock {\em Comm. Math. Phys.}, \textbf{154}:377--401, 1993.

\bibitem{smoller_existence_1993-1}
J.~A. Smoller and A.~G. Wasserman.
\newblock {\em Comm. Math. Phys.}, \textbf{151}:303--325, 1993.

\bibitem{smoller_smooth_1991}
J.~A. Smoller, A.~G. Wasserman, S.-T. Yau, and J.~B. McLeod.
\newblock {\em Comm. Math. Phys.}, \textbf{143}:115--147, 1991.

\bibitem{winstanley_existence_1999}
E.~Winstanley.
\newblock {\em Class. Quant. Grav.}, \textbf{16}:1963--1978, 1999.

\bibitem{bjoraker_monopoles_2000}
J.~Bjoraker and Y.~Hosotani.
\newblock {\em Phys. Rev. D}, \textbf{62}:043513, 2000.

\bibitem{bjoraker_stable_2000}
J.~Bjoraker and Y.~Hosotani.
\newblock {\em Phys. Rev. Lett.}, \textbf{84}:1853--6, 2000.

\bibitem{breitenlohner_non-abelian_2004}
P.~Breitenlohner, D.~Maison, and G.~Lavrelashvili.
\newblock {\em Class. Quant. Grav.}, \textbf{21}:1667, 2004.

\bibitem{winstanley_classical_2009}
E.~Winstanley.
\newblock {\em Lect. Notes Phys.}, \textbf{769}:49--87, 2009.

\bibitem{winstanley_linear_2002}
E.~Winstanley and O.~Sarbach.
\newblock {\em Class. Quant. Grav.}, \textbf{19}:689--724, 2002.

\bibitem{sarbach_linear_2001}
O.~Sarbach and E.~Winstanley.
\newblock {\em Class. Quant. Grav.}, \textbf{18}:2125--2146, 2001.

\bibitem{amann_nodal_1995}
H.~Amann and P.~Quittner.
\newblock {\em J. Math. Phys.}, \textbf{36}:4553--4560, 1995.

\bibitem{wang_invariant_1958}
H.~C. Wang.
\newblock {\em Nagoya Math. J.}, \textbf{13}:1--19, 1958.

\bibitem{yaffe_static_1989}
L.~G. Yaffe.
\newblock {\em Phys. Rev. D}, \textbf{40}:3463--3473, 1989.

\bibitem{baxter_abundant_2008}
J.~E. Baxter, M.~Helbling, and E.~Winstanley.
\newblock {\em Phys. Rev. Lett.}, 100:011301, 2008.

\bibitem{shepherd_characterizing_2012}
B.~L. Shepherd and E.~Winstanley.
\newblock {\em Class. Quant. Grav.}, \textbf{29}:155004, 2012.

\end{thebibliography}
\bibliographystyle{unsrt}
\end{document}